\documentclass[prd,groupedaddress,reprint,floats,nofootinbib,preprintnumbers]{revtex4-2}
\usepackage[T1]{fontenc}
\usepackage{graphicx} 
\usepackage{float}
\usepackage{amsmath}
\usepackage{epstopdf}
\usepackage{booktabs}
\usepackage[normalem]{ulem} % for \sout
\usepackage{xcolor}
\usepackage[breaklinks=true]{hyperref}
\hypersetup{colorlinks=true,urlcolor=magenta,anchorcolor=blue,citecolor=blue,filecolor=blue,linkcolor=magenta,menucolor=blue, linktocpage=true,pdfproducer=medialab}
\usepackage{array,tabularx,epsfig,mathrsfs}
\usepackage{ifthen}
\usepackage{amsfonts}
\usepackage{amssymb}
\usepackage{slashed}
\usepackage{siunitx}
\usepackage{bm} 
\usepackage{diagbox}

\newcommand{\be}{\begin{equation}}
\newcommand{\ee}{\end{equation}}
\newcommand{\ba} {\begin{equation}\begin{aligned}}
\newcommand{\ea} {\end{aligned}\end{equation}}

\newcommand{\sL}{\mathscr{L}}

\newcommand{\derp}{\partial}
\newcommand{\hc}{\text{h.c.}}
\newcommand{\nn}{\nonumber}

\newcommand{\ov}[1]{\overline{#1}}

\newcommand{\TeV}{\ \text{TeV}}
\newcommand{\GeV}{\ \text{GeV}}

\newcommand{\vev}[1]{\langle #1 \rangle}

\newcommand\subsetsim{\mathrel{%
  \ooalign{\raise0.2ex\hbox{$\subset$}\cr\hidewidth\raise-0.8ex\hbox{\scalebox{0.9}{$\sim$}}\hidewidth\cr}}}

\providecommand{\pT}{\ensuremath{p_{\mathrm{T}}}}

\newcommand{\blue}[1]{\color{blue}#1\color{black}}

\newcommand{\magenta}[1]{\color{magenta}#1\color{black}}
\definecolor{darkgreen}{rgb}{0.0, 0.5, 0.0}
\definecolor{lobsterred}{HTML}{e73c2a}

\begin{document}
\preprint{\magenta{IFT-UAM-CSIC-22-152}}
%

%%%%%%%%%%%%%%%%%%%%%%%%%%%%%%%%%%%%%%%%%%%%%%%%%%%%%%%%%%%%%%%
\title{\texorpdfstring{\boldmath\blue{
Probing HNL-ALP couplings at colliders}}%
{Probing HNL-ALP couplings at colliders}}
%%%%%%%%%%%%%%%%%%%%%%%%%%%%%%%%%%%%%%%%%%%%%%%%%%%%%%%%%%%%%%%

\author{Arturo de Giorgi}
\email[]{arturo.degiorgi@uam.es}
\author{Luca Merlo}
\email[]{luca.merlo@uam.es}
\author{Jean-Loup Tastet}
\email[]{jean-loup.tastet@uam.es}
\affiliation{Departamento de F\'isica Te\'orica and Instituto de F\'isica Te\'orica UAM/CSIC,
Universidad Aut\'onoma de Madrid, Cantoblanco, 28049, Madrid, Spain}

%%%%%%%%%%%%%%%%%%%%%%%%%%%%%%%%%%%%%%%%%%%%%%%%%%%%%%%%%%%%%%%
\hfill\draft{ }

\begin{abstract}
Axion-like particles (ALPs) and heavy neutral leptons (HNLs) are both well-motivated extensions of the Standard Model. As ALPs couple to on-shell fermions proportionally to their masses, processes involving both types of particles may give rise, for TeV HNLs, to relevant phenomenology at colliders. In this work, we point out a particularly clean process, whose final state consists of four jets and two charged leptons, and we estimate its current and future sensitivity at the LHC. For on-shell HNLs, i) there is no dependence on the overall scale of the mixing between HNLs and the active neutrinos, making this process sensitive down to the type-I Seesaw line; ii) the signal strength of the process is sizable only as far as the HNL masses are below a few TeVs, contrary to what the proportionality to masses of the ALP couplings may suggest. Although ALPs and HNLs have been mainly studied independently in the literature, considering their interplay may lead to joint limits that are much stronger than those on the individual particles taken separately. This concise study paves the way for similar searches at colliders and other experiments. 
\end{abstract}

\maketitle
\section{Introduction}
\label{sec:introduction}

Axions and Axion-like particles (ALPs) experienced a revival of interest in the last years. From a more theoretical point of view, several studies appeared on axion constructions that extend the Peccei-Quinn model~\cite{Peccei:1977hh,Weinberg:1977ma,Wilczek:1977pj} and the original invisible axion models~\cite{Zhitnitsky:1980tq,Dine:1981rt,Kim:1979if,Shifman:1979if}, investigating possible axion flavour dynamics~\cite{Ema:2016ops,Calibbi:2016hwq,Arias-Aragon:2017eww,Arias-Aragon:2020qip,Arias-Aragon:2022ats}, whose pioneering idea goes back to Ref.~\cite{Wilczek:1982rv}, or analysing the axion presence in composite Higgs frameworks~\cite{Merlo:2017sun,Brivio:2017sdm,Alonso-Gonzalez:2018vpc}, or focusing on studying and enlarging the classical QCD axion parameter space~\cite{DiLuzio:2016sbl,DiLuzio:2017pfr,Gaillard:2018xgk,Hook:2019qoh,DiLuzio:2020wdo,DiLuzio:2020oah,DiLuzio:2021pxd,DiLuzio:2021gos}, showing that QCD axions may be much lighter or much heavier than in the classical scenario. Given the latter, we will adopt in the following the wording ALPs to refer to both QCD axions and to generic Goldstone bosons whose mass is not originated by non-perturbative QCD effects.

On the experimental side, the efforts pointed towards the construction of a consistent effective description of ALPs~\cite{Brivio:2017ije,Alonso-Alvarez:2018irt,Gavela:2019wzg,Chala:2020wvs,Bonilla:2021ufe,Bauer:2021wjo,Arias-Aragon:2022byr,Arias-Aragon:2022iwl}, beyond the seminal study in Ref.~\cite{Georgi:1986df}, in order to investigate possible signals at low-energy facilities~\cite{Izaguirre:2016dfi,Merlo:2019anv,Bauer:2019gfk,Bauer:2020jbp,Bauer:2021mvw,Guerrera:2021yss,Gallo:2021ame,Bonilla:2022qgm,Bonilla:2022vtn,deGiorgi:2022vup,Guerrera:2022ykl}, colliders~\cite{Jaeckel:2012yz,Mimasu:2014nea,Jaeckel:2015jla,Alves:2016koo,Knapen:2016moh,Brivio:2017ije,Bauer:2017nlg,Mariotti:2017vtv,Bauer:2017ris,Baldenegro:2018hng,Craig:2018kne,Bauer:2018uxu,Gavela:2019cmq,Haghighat:2020nuh,Wang:2021uyb,Bonilla:2022pxu,Ghebretinsaea:2022djg,Vileta:2022jou} and in non-terrestrial environments~\cite{DEramo:2018vss,Arias-Aragon:2020qtn,Arias-Aragon:2020shv,DEramo:2022nvb}. Most of the analyses performed on collider data and from astrophysics deal with an on-shell ALP in the initial or final state of the considered processes, such as mono-$Z$, mono-$W$, and mono-Higgs searches or stellar cooling, and the associated bounds strongly depend on the ALP mass. The common aspect is that the associated Feynman diagram has only one single ALP coupling insertion and therefore the corresponding process amplitude is suppressed by only one power of the ALP characteristic scale $f_a$. On the other hand, the lack of observation of any tree-level flavour-changing process mediated by neutral currents leads to very strong bounds on ALP couplings, for example from meson oscillations and (semileptonic) meson decays. In this case, and in those few non-resonant ALP searches at colliders, the associated amplitude is suppressed by $f_a^2$ as the ALP appears as an internal line of the Feynman diagram.

Within the Standard Model (SM) framework with the addition of an ALP, it is always possible to express the ALP couplings to fermions in the chirality flip basis where the proportionality to the corresponding fermion mass becomes manifest. This implies that the heavier the fermion, the more enhanced is the corresponding coupling to the ALP. From here the idea is to study the implications of ALP couplings to Heavy Neutral Leptons (HNLs), typically introduced in Seesaw mechanisms~\cite{Minkowski:1977sc,Mohapatra:1979ia,Yanagida:1979as,Glashow:1979nm,Gell-Mann:1979vob,Schechter:1980gr,Schechter:1981cv,Mohapatra:1986bd,Bernabeu:1987gr,Malinsky:2005bi} and with masses larger than the Electroweak (EW) scale. The idea of coupling an ALP to HNLs has not received much attention in the literature and only a few studies have appeared~\cite{PhysRevD.96.115035,Alves:2019xpc,Dekker:2021bos,Gola:2021abm,Arias-Aragon:2021rve}.

Without entering into the details of any specific model of neutrino masses, HNLs are states composed of a mixture between the neutral components of the SM lepton Electroweak (EW) doublets and sterile leptons, singlets under the SM gauge symmetries: the matrix containing the mixings is typically labeled as $\bm{\Theta}$ and its entries are strongly bounded from global analyses on non-unitarity effects of the PMNS mixing matrix~\cite{Langacker:1988ur,Antusch:2006vwa,Fernandez-Martinez:2007iaa,Antusch:2014woa,Fernandez-Martinez:2016lgt,Arguelles:2022xxa}.
There exists a strong experimental program dedicated to searching for HNLs (see e.g.\ Refs.~\cite{Beacham:2019nyx,Agrawal:2021dbo,Abdullahi:2022jlv} for detailed reviews of existing and proposed searches), including at the CMS~\cite{CMS:2022rqc,CMS:2022fut,CMS:2018iaf,CMS:2018jxx,CMS:2016aro,CMS:2015qur,CMS:2012wqj}, ATLAS~\cite{ATLAS:2022atq,ATLAS:2019kpx,ATLAS:2015gtp,ATLAS:2011izm} and LHCb~\cite{LHCb:2014osd} experiments at the LHC, where HNLs have been probed over a wide range of masses, in both the prompt and displaced regimes.
As one of the main \emph{renormalizable portals}~\cite{Asaka:2005an,Asaka:2005pn,Patt:2006fw,Cerdeno:2006ha,Batell:2009di} to new physics, HNLs are also a prime target for current and future searches at displaced detectors~\cite{FASER:2018bac,Kling:2018wct,SHiP:2020sos,Curtin:2018mvb,Aielli:2019ivi,Gligorov:2018vkc,Dercks:2018wum,Bauer:2019vqk,Hirsch:2020klk,Feng:2022inv,Anchordoqui:2021ghd,Cerci:2021nlb}, extracted beamlines~\cite{NA62:2017qcd,NA62:2020mcv,NA62:2021bji,Tastet:2020tzh,E949:2014gsn,Baldini:2021hfw,Bonivento:2013jag,SHiP:2015vad,Alekhin:2015byh,SHiP:2018xqw,Tastet:2019nqj,SHiP:2021nfo,DUNE:2021tad,Ballett:2019bgd,Berryman:2019dme,Krasnov:2019kdc,Coloma:2020lgy,Breitbach:2021gvv}, colliders~\cite{Belle:2022tfo,Blondel:2014bra,Antusch:2016ejd,Blondel:2022qqo,Shen:2022ffi,Mekala:2022cmm,Pascoli:2018heg,Ruiz:2017yyf,Antusch:2019eiz,Rodejohann:2010jh,Black:2022cth,Bose:2022obr,Cvetic:1999fk,Almeida:2000qd}, and cLFV experiments~\cite{Calderon:2022alb,Crivellin:2022cve}.
A number of reinterpretations have furthermore appeared, either to correct~\cite{Shuve:2016muy} the reported limits or to extend them~\cite{Abada:2018sfh,Tastet:2021vwp,Abada:2022wvh} to non-minimal scenarios consistent with the observed lightness of neutrino masses.

\begin{figure}[ht]
    \centering
    \includegraphics[width=0.9\linewidth]{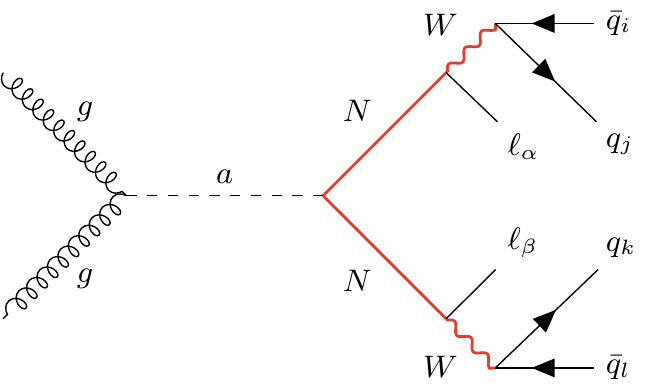}
    \caption{\em Feynman diagram of the JALZ-topology relevant for the study. The on-shell particles (besides final and initial states) are shown in {\color{lobsterred}red}.}
    \label{fig:diagram}
\end{figure}

In this letter, we identify a promising clean process that may shed light on the ALP couplings to HNLs and we study its sensitivity prospects at present and future colliders. The corresponding lobster-like Feynman diagram is shown in Fig.~\ref{fig:diagram}. The ALP is produced off-shell through gluon fusion, which largely dominates over the other production mechanisms. 
The ALP then decays into two on-shell HNLs $N$, both of which subsequently decay into a charged lepton $\ell$ and an on-shell $W$ gauge boson, whose final decay is into jets. The resulting final state consists of four jets and two charged leptons with same or opposite signs (j4$\ell$2), which we refer to as the JALZ topology\footnote{Such topology has been previously considered in Ref.~\cite{FileviezPerez:2009hdc} in the context of a $Z'$-portal to HNLs, and in Ref.~\cite{Biggio:2011ja} for the type-III Seesaw.}. For HNLs with mass $M_N \gg M_W$, each $W$ would be significantly boosted, resulting in an overlap between its two outgoing jets and making it harder to reconstruct the event unless a dedicated large-radius reconstruction~\cite{ATLAS:2020gwe} is used (see App.~\ref{app:efficiencies}).
 
This specific setup has the following advantages:
\begin{itemize}
\item[-] the whole process is proportional to the HNL mass, thus the signal strength is sizeable;
\item[-] it is possible to adopt the narrow width approximation and then the branching ratios of the HNL decays do not depend on the overall scale of the mixing $\bm{\Theta}$, typically very small;
\item[-] the on-shellness of the two HNLs and the two $W$'s, combined with the absence of light neutrinos, leads to a fully reconstructible final state with four \emph{simultaneous} mass peaks. Not only we do not expect any Standard Model background with such a kinematic structure, but it could also significantly suppress the background from pileup jets that a process with four jets would otherwise suffer from.
\end{itemize}

With respect to other collider analyses, the JALZ topology is almost independent of the ALP mass and it is doubly suppressed by the ALP characteristic scale $f_a$, whose effect is however mitigated by the enhancement of the HNL mass. One may expect that the larger the HNL mass, the bigger would be the signal strength of the process, but this is not the case: due to the parton distribution functions of the colliding protons, the signal strength drops very fast for masses larger than $\sim \mathcal{O}(1.5)$ TeV. 

In what follows, we will first introduce the parts of the ALP and HNL Lagrangian relevant to discuss the JALZ topology, then we will present the corresponding numerical analysis, and we will conclude with the sensitivity to the ALP-HNL coupling at colliders for different integrated luminosities as the main result of the letter.

\section{The ALP-HNL effective Lagrangian}
\label{sec:lagrangian}

Without entering into the details of a specific high energy theory, we will assume that a global symmetry, alike the Peccei-Quinn one, gets spontaneously broken at some large scale, giving rise to an ALP that couples to the SM spectrum, enlarged by $n$ right-handed (RH) neutrinos, with $n\geq2$. The latter provides lepton number violation that leads to masses for the active neutrinos at low energies: without focusing on a specific Seesaw mechanism, we will simply assume a minimal set of terms in the Lagrangian that describe the effective interactions at a scale where the HNLs are dynamical. 

The effective Lagrangian that we will adopt for our analysis can be written as the sum of different terms:
\be
\sL^\text{eff}=\sL_\text{SM}+\sL_\text{HNL}+\sL_a\,,
\label{eq:defLeff}
\ee
where the first term is the SM Lagrangian, the second contains the interactions of the HNL with the SM spectrum and the last describes the ALP couplings. More specifically, 
\be
%\sL_\text{HNL}=i\ov{N'_R}\slashed{D}N'_R-\ov{L'_L}\bm{Y}_N\widetilde H N_R'-\dfrac12\ov{N_R^{\prime c}}\bm{M}_N N'_R+\hc\,,
%\sL_\text{HNL}=i\ov{N'_R}\slashed{D}N'_R- \left( \ov{L'_L}\bm{Y}_N\widetilde H N_R' + \dfrac12\ov{N_R^{\prime c}}\bm{M}_N N'_R+\hc\right),
\sL_\text{HNL}=i\ov{N'_R}\slashed{D}N'_R- \left( \ov{L'_L} \widetilde H \bm{Y}_N N_R' + \dfrac12\ov{N_R^{\prime c}}\bm{M}_N N'_R+\hc\right),
\label{eq:LHNL}
\ee
where $L'_L$ and $H$ stand for the EW doublets of leptons and Higgs respectively, with $\widetilde H=i
\sigma_2 H^\ast$, and $N'_R$ are the RH neutrinos. The primes symbolise a generic flavour basis, where $\bm{M}_N$ and $\bm{Y}_N$ are $n\times n$ (symmetric) and $3\times n$ matrices in the flavour space, respectively. After the EW symmetry breaking, the active neutrinos acquire small masses: in the canonical type-I Seesaw mechanism~\cite{Minkowski:1977sc,Mohapatra:1979ia,Yanagida:1979as,Glashow:1979nm,Gell-Mann:1979vob,Schechter:1980gr,Schechter:1981cv} their lightness results from the hierarchy between the Dirac and Majorana mass terms, while in the low-scale Seesaw variations~\cite{Mohapatra:1986bd,Bernabeu:1987gr,Malinsky:2005bi,Asaka:2005an,Asaka:2005pn} it follows from an approximately conserved lepton number~\cite{Kersten:2007vk,Shaposhnikov:2006nn,Gluza:2002vs,Moffat:2017feq,Drewes:2019byd,Fernandez-Martinez:2022gsu} and the HNL masses can be close to the EW scale. Once the Higgs develops its vacuum expectation value (vev), $\vev{H}=v/\sqrt2$ with $v=246\GeV$, the neutral leptons mix, and the active neutrino mass term reads
\be
\sL_\nu\supset-\dfrac12\ov{\nu^{\prime\prime}_L}\,\bm{m}_\nu\,\nu_L^{\prime\prime c}+\hc\,,
\ee
where in the Seesaw limit $v\ll||\bm{M}_N||$,
\be
\bm{m}_\nu=-\bm{\Theta}\, \bm{M}_N\,\bm{\Theta}^T
\label{eq:seesaw}
\ee
with the mixing matrix $\bm{\Theta}$ defined as
\be
\nu'_L\equiv\nu^{\prime\prime}_L+\bm{\Theta} N_R^{\prime\prime c}\,,\qquad
\bm{\Theta}=\dfrac{v}{\sqrt2} \bm{Y}_N\,\bm{M}_N^{-1}\,.
\label{defTheta}
\ee
The HNL mass matrix can be identified with $\bm{M}_N$, neglecting small corrections proportional to the active neutrino mass matrix. Due to the mixing in Eq.~\eqref{defTheta}, the HNLs acquire couplings proportional to $\bm\Theta$ with the EW gauge and scalar bosons and with the other leptons.

The fermion mass basis is achieved by diagonalising the mass matrix of the charged leptons, $\bm{m}_\nu$, and $\bm{M}_N$, 
\be
\widehat{\bm{m}}_\nu=\bm{U}^\dag_\nu\,\bm{m}_\nu\, \bm{U}^\ast_\nu\,,\qquad
\widehat{\bm{M}}_N=\bm{U}^T_N\,\bm{M}_N\, \bm{U}_N\,,
\ee
and correspondingly the PMNS mixing matrix appears in the $W$--lepton currents. The associated mass eigenstates will be denoted as $\nu_L$ and $N_R$. As already mentioned, depending on the specific Seesaw realisation, $N_R$ can be heavy ($\sim 10^{15}\,\si{GeV}$) or very light ($\sim \si{MeV}$). In the latter case, its phenomenological study is in general very promising in both direct and indirect searches.
In the absence of any cancellation between the elements of $\mathbf{\Theta}$ in Eq.~\eqref{eq:seesaw}, the light neutrino mass matrix is of order\footnote{The precise value depends on the number of HNLs, the neutrino mass ordering and the specific elements of $\bm{\Theta}$.} $||\bm{m}_{\nu}|| \sim ||\bm{M}_N|| \cdot ||\bm{\Theta}||^2$, and conversely
\be
    ||\bm{\Theta}||^2 \sim \frac{||\bm{m}_{\nu}||}{||\bm{M}_N||}.
\ee
Plotted as a function of the HNL mass(es) and taking $||\bm{m}_{\nu}|| \sim \SI{50}{meV}$, this ``Seesaw line'' represents the naive expectation for the mixing matrix, in the absence of an approximate lepton number symmetry.

The last term in Eq.~\eqref{eq:defLeff} can be written as
\be
\begin{aligned}
\sL_a=&\dfrac{1}{2}\derp_\mu a\,\derp^\mu a-\dfrac12m_a^2\,a^2-\dfrac{a}{f_a}\sum_{X}c_{aXX}\,X^{\mu\nu}\widetilde X_{\mu\nu}+\\
% &-\dfrac{\derp_\mu a}{f_a}\left(\ov{L'_L}c'_L\gamma^\mu L'_L+\ov{\ell'_R}c'_\ell\gamma^\mu \ell'_R+\ov{N'_R}c'_N\gamma^\mu N'_R\right)\,,
&-\dfrac{\derp_\mu a}{f_a}\sum_{\psi'}\ov{\psi'}\bm{c}'_\psi\gamma^\mu \psi'
\end{aligned}
\label{eq:LALP}
\ee
where $X^{\mu\nu}$ is the field strength tensor of the SM gauge bosons and $\widetilde X_{\mu\nu}$ its dual, $m_a$ is the generic ALP mass (that does not need to originate from non-perturbative QCD effects) and $\psi'$ stands for any fermion field in the spectrum in the initial flavour basis, with $\bm{c}'_\psi$ the corresponding ALP-fermion coupling being a matrix in the flavour space. In the fermion mass basis, the ALP-HNL coupling contained in the second line in the previous expression reads
\be
\sL_a\supset-\dfrac{\derp_\mu a}{f_a}\,\ov{N_R}\,\gamma^\mu\,\bm{c}_N\, N_R
\quad \text{with}\quad
\bm{c}_N\equiv \bm{U}_N^\dag \bm{c}'_{N_R} \bm{U}_N\,.
\ee
The latter coupling and the anomalous ALP-gluon coupling in Eq.~\eqref{eq:LALP} are the basic ingredients necessary to discuss the JALZ process. The ALP couplings to charged fermions and active neutrinos have already been studied in the literature. The ALP couplings to one active neutrino and one HNL have not been investigated, but they are proportional to $\bm{\Theta}$ and therefore the associated signals are suppressed and harder to see at present and near-future colliders.

%^^^^^^^^^^^^^^^^^^^^^^^^
\section{The analysis}

In the rest of the letter, we will discuss the phenomenology of the lightest HNL, dubbed $N$. Consistently, its mass will be denoted by $M_N$ (not boldface), its couplings with the other leptons will only include the lepton family index and will be denoted by $\Theta_\alpha$, while the one with the ALP will be indicated as $c_N$ (not boldface).

\paragraph{On-shellness.}
The on-shellness of the HNLs allows separating their production and decay using the narrow-width approximation. For a given combination of lepton flavours $\alpha,\beta \in e,\mu,\tau$ with any combination of charges:
\be
\begin{split}
%\sigma(pp \to a^* \to &(N \to \ell_{\alpha} q_i q_j) (N \to \ell_{\beta} q_k q_l)) = \\
%=&\frac{2}{\mathcal{S}} \times \sigma(pp \to a^* \to NN) \times\\ 
%&\times \mathrm{Br}(N \to \ell_{\alpha} q_i q_j)\times\mathrm{Br}(N \to \ell_{\beta} q_k q_l)
\sigma_{\alpha\beta} \equiv \sigma(pp \to a^* \to &(N \to \ell_{\alpha} \bar{q}_i q_j) (N \to \ell_{\beta} q_k \bar{q}_l)) = \\
=&\frac{2}{\mathcal{S}} \times \sigma(pp \to a^* \to NN) \times\\ 
&\times \mathrm{Br}(N \to \ell_{\alpha} \bar{q}_i q_j)\times\mathrm{Br}(N \to \ell_{\beta} q_k \bar{q}_l)
\end{split}
\label{eq:NWA}
\ee
where $\mathcal{S}$ is a symmetry factor equal to $2$ if both HNL decays are identical and to $1$ otherwise, and the $2$ in the numerator is a combinatorial factor accounting for the presence of two HNLs.
For a given choice of lepton flavours, $\alpha$ and $\beta$, Eq.~\eqref{eq:NWA} shows that the cross-section scales as follows with the various parameters (excluding the HNL mass and neglecting the charged lepton masses):
\begin{equation}
    \sigma_{\alpha\beta} \propto \frac{c_{agg}^2 c_N^2}{f_a^4} \frac{|{\Theta}_{\alpha}|^2 |{\Theta}_{\beta}|^2}{\Gamma_N^2}
    \propto \frac{c_{agg}^2 c_N^2}{f_a^4} \frac{|\Theta_{\alpha}|^2}{|\Theta|^2} \frac{|\Theta_{\beta}|^2}{|\Theta|^2}
    \label{eq:scaling}
\end{equation}
where we have defined $|{\Theta}|^2 = \sum_{\alpha=e,\mu,\tau} |{\Theta}_{\alpha}|^2$ and used the proportionality of the total HNL width to $|{\Theta}|^2$. This shows that the cross-section is independent of the overall scale of the HNL mixings.

If we were to sum the cross-section over all three lepton flavours, the dependence on the mixing pattern $|\Theta_\alpha|^2/|\Theta|^2$ would disappear too (up to small finite-mass corrections). However, in practice, different lepton flavours come with different detection efficiencies, resulting in a residual dependence on the mixing pattern. Nevertheless, we should expect the sensitivity of the process to depend only weakly on the ratio of the three mixings.
If, however, we were to consider separately the processes involving different combinations of charged lepton flavours, we could in principle constrain or even measure the relative values of the $\bm{\Theta}$ entries.

\paragraph{HNL lifetime.}

The expected number of signal events can be expressed as:
\begin{equation}
    N_{\mathrm{events}} = L_{\mathrm{int}} \sigma \epsilon
\end{equation}
with $L_{\mathrm{int}}$ denoting the integrated luminosity, $\sigma$ the cross-section of the process, and $\epsilon$ the signal efficiency of the process.
Despite the fact that the dependence on the overall $|\Theta|^2$ scale cancels out, as shown by Eq.~\eqref{eq:scaling}, if the HNL is long-lived, it can fly out of the detector, lowering the efficiency and reintroducing a dependence on $|\Theta|^2$.
In particular, when the HNL lifetime becomes much larger than the size of the experiment $r_{\mathrm{exp}}$, the efficiency (for each HNL) will have to include the probability of decaying within the detector, which asymptotically goes as $\Gamma_N r_{\mathrm{exp}} / \gamma_N \propto |{\Theta}|^2 / \gamma_N$, with $\gamma_N$ the HNL boost that only depends on $M_N$. This would make the signal strength proportional to $|\Theta|^4$.
App.~\ref{app:formulae} discusses the HNL lifetime in more detail and shows that for models of interest ($|{\Theta}|^2$ at or above the Seesaw line) and at the LHC, the HNLs can safely be considered as decaying promptly (see Fig.~\ref{fig:lifetime}), and therefore the signal strength is independent of the overall scale of the HNL mixings.

\paragraph{Other channels.}
The topology presented in Sect.~\ref{sec:introduction} can lead to a number of different signatures, depending on the HNL decays and on the subsequent decays of the resulting on-shell bosons.
When $M_N \gg M_h$, each HNL decays to $W:Z:h$ in proportions $\frac{1}{2}:\frac{1}{4}:\frac{1}{4}$, and each $W$ decays hadronically (leptonically) about $2/3$ ($1/3$) of the time.
Out of all the final states, only the JALZ topology is in principle fully reconstructible, with no neutrinos in the final state. It corresponds to the lobster-like diagram shown in Fig.~\ref{fig:diagram}, and will be the focus of this study. Because each HNL takes the fully reconstructible decay channel $\ell jj$ about $1/3$ of the time, this final state is realised in about $1/9$ of the $NN$ pairs produced through the coupling of HNLs to the ALP. This is in agreement with our numerical estimates, up to finite-mass effects that correct the decay widths when $M_N$ is near the $\mathcal{O}(100\,\mathrm{GeV})$ scale.
Another $\approx 80\%$ of the final states resulting from the decay of the $NN$ pair contain at least one neutrino --- and will therefore lead to a signature with some missing transverse energy (MET). Half of them contain one fully reconstructible HNL decay, while the other half contains MET in both decays. If one can deal with the increased background resulting from not being able to reconstruct the mass of both HNLs, these could in principle be interesting search channels.
In fact, similar topologies (mediated by on-shell fermions and vector bosons, and with off-shell gauge bosons instead of the ALP) have already been searched for at the LHC in the context of the type-III Seesaw and vector-like leptons, see e.g.\ Refs.~\cite{ATLAS:2015fma,ATLAS:2015qoy,CMS:2019hsm,CMS:2019lwf,ATLAS:2020wop,ATLAS:2022yhd}, although in the models they consider, the production mechanism is not gluon fusion but Drell-Yan.
Tab.~\ref{tab:channels} summarises the relative frequencies of all the final states, classified in terms of whether each HNL decay is reconstructible, partially reconstructible (MET), or invisible.

\begin{table}[tbh]
\centering
\begin{tabular}{|l|*{3}{c}|}
\hline
&&&\\[-3mm]
\backslashbox{First $N$}{Second $N$} & Reconstructible & MET & Invisible\\
\hline
&&&\\[-3mm]
Reconstructible & $10\%$ &  &  \\[1mm]
MET & $40\%$ & $40\%$ &   \\[1mm]
Invisible & $3\%$ & $6\%$ & $0.2\%$  \\
\hline
\end{tabular}
 \caption{Approximate relative frequencies of the various final states, classified according to the degree to which each HNL decay can be reconstructed, for $M_N \gg M_h$. ``First'' and ``Second'' denote the two HNLs produced in the ALP decay, in no particular order since they are indistinguishable.}
\label{tab:channels}
\end{table}

\begin{figure}[t]
    \centering
    \includegraphics[width=.9\linewidth]{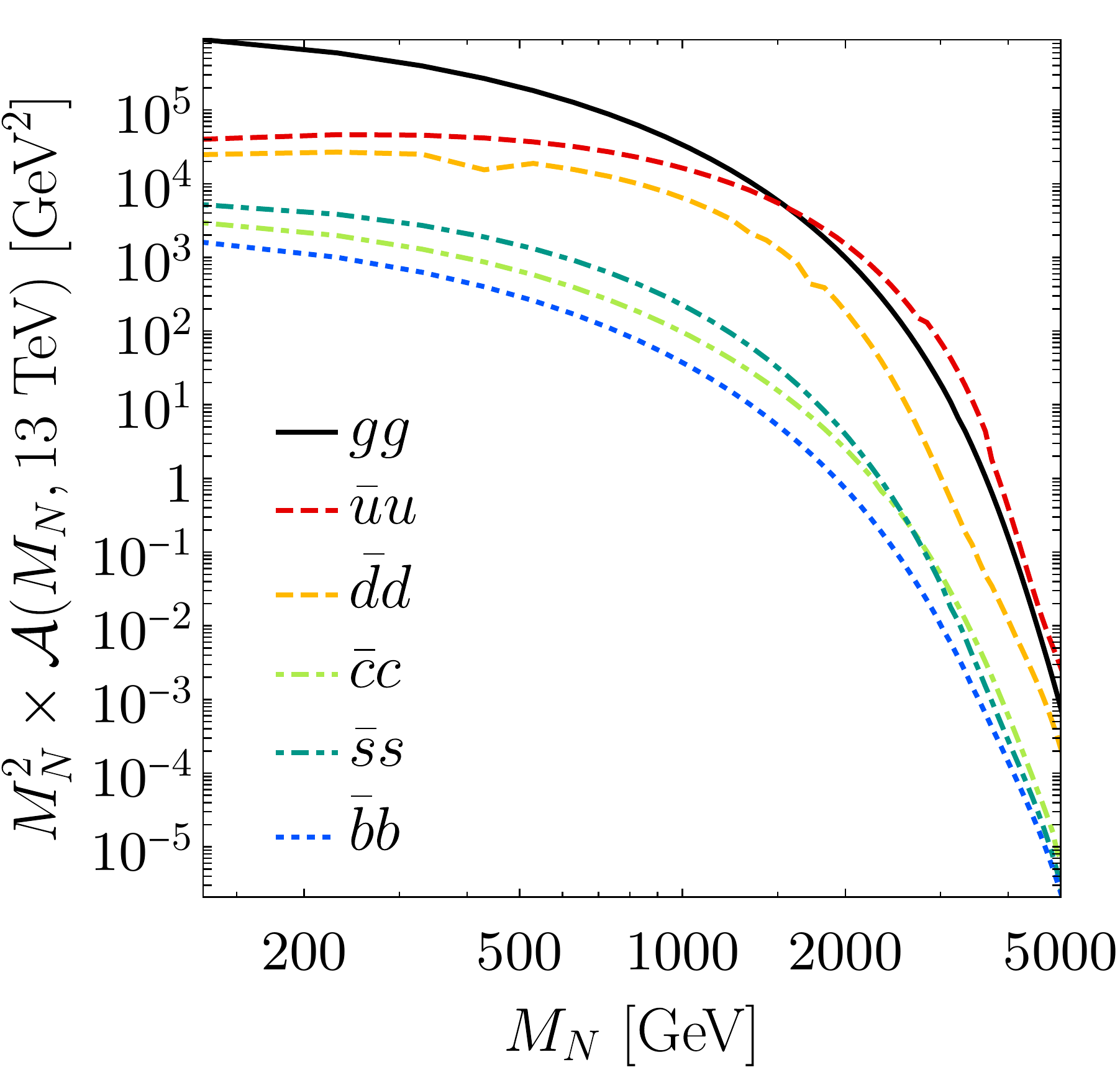}
    \caption{\em Scaling of the PDFs area times $M_N^2$ for different partons involved in the HNLs production as a function of $M_N$.
    }
    \label{fig:area}
\end{figure}

%^^^^^^^^^^^
\paragraph{Impact of partonic effects.}
The effect of Parton Distribution Functions (PDFs) is very relevant. Not all the partons within the proton have the same energy. Denoting by $x_{1,2}$ the momentum fractions of the two partons, the total energy at their disposal is $\hat{s}\equiv x_1x_2 s$. In order to produce two HNLs, the kinematic condition $\sqrt{\hat{s}}\geq 2 M_N$ must hold. The area subtended by the PDFs, and that has to be convoluted with the partonic cross-section, gets reduced considerately at high momentum fraction. Therefore, as the mass of the HNL grows, the kinematic condition has little space to be satisfied, thus suppressing the final cross-section. To get a rough estimate of such an effect, we compute the area of the PDFs that must be convoluted with the partonic cross-section
\begin{equation}
\mathcal{A}_{g,q}(M_N, \sqrt{s})\equiv \iint_{\sqrt{\hat{s}}\geq 2M_N}\text{d}x_1\text{d}x_2 f_{g,q}(x_1)f_{g,\Bar{q}}
(x_2)\,,
\end{equation}
where $f_i(x)$ are the PDFs. The result for $\sqrt{s}=13$ TeV is shown in Fig.~\ref{fig:area}. In this Figure, we have used the central values from the PDFs \textit{MSTW 2008}~\cite{Martin:2009iq}.
We have conveniently multiplied $\mathcal{A}$ by $M_N^2$ as this is the naively expected enhancement given by the ALP vertex.

%%^^^^^^^^^
\paragraph{Monte-Carlo event generation.}
The cross-sections are computed numerically using the \texttt{MadGraph5\_aMC@NLO} (v3.4.1) Monte-Carlo event generator~\cite{Alwall:2014hca}.
The Lagrangian defined in Sect.~\ref{sec:lagrangian} is implemented in the \texttt{Mathematica}~\cite{Mathematica} package \texttt{FeynRules} (v2.3.41)~\cite{Alloul:2013bka}, taking some inspiration from the \texttt{HeavyN}~\cite{Alva:2014gxa,Degrande:2016aje} and \texttt{ALP\_linear}~\cite{Brivio:2017ije} models.
Samples are generated for a number of HNL masses within the range $[130,5000]\,\mathrm{GeV}$.
After numerically verifying that gluon fusion $gg \to a^*$ is the dominant production mechanism throughout this mass range, we neglect the remaining production modes. We use the PDF set \texttt{NNPDF40\_nnlo\_as\_01180} from the NNPDF collaboration~\cite{NNPDF:2021njg} through the \texttt{LHAPDF6} interface~\cite{Buckley:2014ana}. The ALP mass is set to a non-zero value of $1\,\mathrm{GeV}$ to avoid numerical stability issues, but is shown to have no effect on the cross-sections. We then parse the results file to aggregate the cross-sections for the various signatures of interest, using the scaling properties \eqref{eq:scaling} as needed to interpolate between the various HNL mixing patterns (see Refs.~\cite{FIPs2022Report} or \cite[][sec.~3]{Tastet:2021vwp} for a description of this method).
%%^^^^^^^^^

\section{The Projected Sensitivity}
The JALZ topology is selected by requiring no neutrinos in the final state, and the projected single-event sensitivity of an LHC experiment is computed for various luminosity targets. The underlying assumption is that simultaneously reconstructing the two HNL masses, each from the momenta of a lepton and of a large-radius, $W$-tagged jet~\cite{ATLAS:2020gwe}, suppresses the background down to a negligible level (the correct combination of lepton and jet being selected by requiring the reconstructed HNL masses to agree).
Using the scaling law in Eq.~\eqref{eq:scaling}, the projected sensitivity is then expressed in terms of $\sqrt{c_{agg} c_N} / f_a$ and the HNL mass.

Although summing Eq.~\eqref{eq:scaling} over flavours leads to a cross-section that is completely independent of HNL parameters, in practice efficiencies differ for each flavour combination. This is particularly true for the tau lepton, whose leptonic modes account for $35\%$ of decays and do not allow to identify the tau as such at the LHC. They would lead to a final state similar to the JALZ topology, but with missing transverse energy that may impede the reconstruction of the HNL mass. In what follows we will therefore focus on hadronically decaying taus $\tau_{\mathrm{h}}$, and we assume a further reconstruction/identification efficiency\footnote{The reconstruction and identification efficiencies of tau leptons crucially depend on the chosen working point, i.e.\ on the required level of background rejection (see e.g.\ Ref.~\cite{ATLAS:2020azv}). The favourable background conditions of the JALZ topology should allow for a high efficiency, hence the chosen value of $80\%$.} of $80\%$ for each tau, leading to an overall efficiency of $\epsilon_{\tau} \approx 0.5$.

\begin{figure}[t]
    \centering    \includegraphics[width=.95\linewidth]{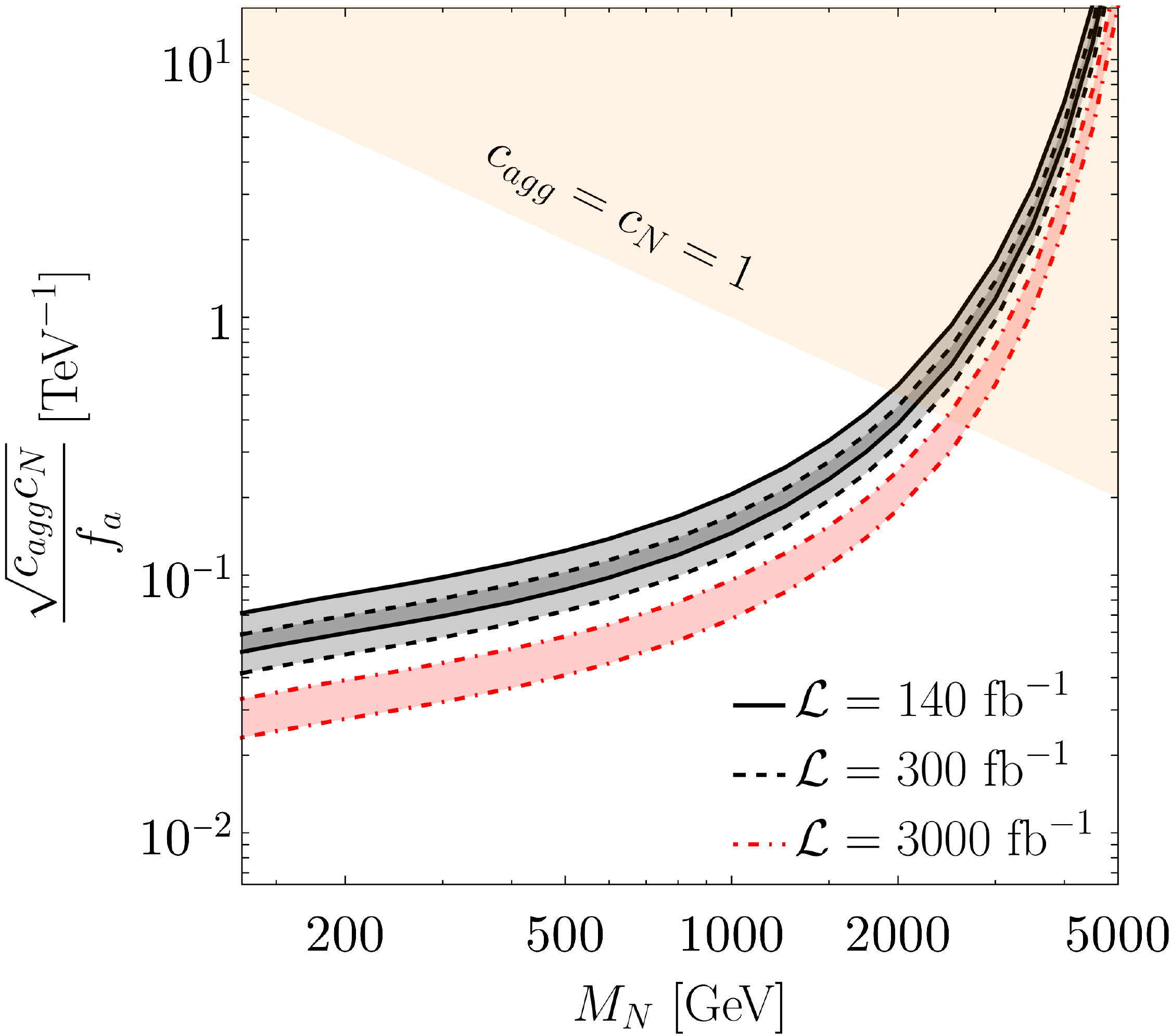}
    \caption{\em Sensitivity to NP couplings as a function of the HNL mass at LHC for different integrated luminosities at $\sqrt{s}=13$ TeV, assuming perfect efficiency for electrons, muons, and jets. Above the lines, a number of events $N_\text{events}>1$ is expected. For each colored band, the bottom line corresponds to the HNL mixing only with $e$ and $\mu$ (best sensitivity), and the top one to mixing only with $\tau$ (worst sensitivity). 
    The ratio of the two lines corresponds to $\sqrt{\epsilon_{\tau}}$.
    Any intermediate case is expected to lie within the band.
    Finally, the orange region shows where $f_a\leq M_N$ for $c_{gg}=c_N=1$, where the theory is expected to become unreliable.
    }
    \label{fig:parameter-space}
\end{figure}

The final result of the letter is in Fig.~\ref{fig:parameter-space} where we show the sensitivity to the ALP couplings as a function of the HNL mass at the LHC for different integrated luminosities. The thickness of the sensitivity bands is defined considering HNL mixing with only electrons and muons (lower line), or only with the taus (upper line). Because these limits do not take into account signal efficiencies other than the tau's, we expect them to be slightly optimistic. In App.~\ref{app:efficiencies}, we apply a simple event selection, discuss the most important cuts at both low and high masses, and argue for the use of large-radius jet reconstruction to search for the JALZ topology. \\

\section{Concluding remarks}
The existence in Nature of ALPs from one side and of HNLs from the other is very well motivated. In this letter, we assume that both types of particles exist and interact with each other, and we point out an efficient discovery channel at colliders, whose final state consists of four jets and two charged leptons. When a non-resonant ALP produces two on-shell HNLs, with masses above the electroweak scale, the signal strength at present and near-future colliders is independent of the overall scale of the HNL mixing with the active neutrinos: this is particularly appealing as the elements of the $\bf\Theta$ matrix are constrained to be tiny. However, the signal strength still depends on the relative values of the $\bf\Theta$ entries. In case of not distinguishing between the charged leptons in the final state, the dependence on $\bf\Theta$ would completely disappear in the cross section, but a mild dependence in the signal strength remains due to the different efficiencies. On the contrary, in the event of an observation of this process, distinguishing between the charged leptons in the final state would provide constraints on $|\Theta_\alpha|/|\Theta_\beta|$ combinations.

Moreover, the analysis is independent of the ALP mass, thus applying to generic Goldstone bosons as well as to QCD axions, with $f_a$ not too much larger than tens of TeV (a modification of the traditional QCD axion parameter space has been recently discussed in the literature~\cite{DiLuzio:2016sbl,DiLuzio:2017pfr,Gaillard:2018xgk,Hook:2019qoh,DiLuzio:2020wdo,DiLuzio:2020oah,DiLuzio:2021pxd,DiLuzio:2021gos}).

The standard lore for ALP-fermion couplings is that they are proportional to the corresponding fermion mass, thus naively implying that the larger the mass, the higher the signal strength of a process induced by such a coupling. However, we showed that the impact of the PDFs is very much relevant, leading to a drop of the signal strength (and of the sensitivity to the couplings) for HNL masses larger than a couple of TeV for LHC data with $\sqrt{s}=13\TeV$.

This study represents a first example of how a dedicated analysis of the interactions between ALPs and HNLs could shed light on the nature of these elusive particles. In particular, we showed that their interplay may lead to joint limits that are much stronger than those on the individual particles taken separately. It also paves the way for model building constructions and for additional searches at colliders and other types of experiments.

\begin{acknowledgments}
The authors acknowledge partial financial support by the Spanish Research Agency (Agencia Estatal de Investigaci\'on) through the grant IFT Centro de Excelencia Severo Ochoa No CEX2020-001007-S and by the grant PID2019-108892RB-I00 funded by MCIN/AEI/ 10.13039/501100011033, by the European Union's Horizon 2020 research and innovation programme under the Marie Sk\l odowska-Curie grant agreement No 860881-HIDDeN.
JLT acknowledges partial support from the grant Juan de la Cierva FJC2021-047666-I funded by MCIN/AEI/10.13039/501100011033 and by the European Union ``NextGenerationEU''/PRTR.
\end{acknowledgments}

%%%%%
%%%%%%%%%%%%%%%%%%%%%%%%%  Appendix    %%%%%%%%%%%%%%%%%%%%%%%%
\begin{appendix}
\section{HNL Lifetime}
    \label{app:formulae}
    For HNLs heavier than the Higgs, the decays to on-shell $W$'s, $Z$'s, and $h$'s dominate, with rates~\cite{Gronau:1984ct}:
\begin{align}
    \Gamma(N \to W\ell_\alpha) &\approx \frac{G_F M_N^3}{4\pi\sqrt{2}}  |\Theta_\alpha|^2 \left( 1 - \frac{M_W^2}{M_N^2} \right)^2 \left( 1 + \frac{2M_W^2}{M_N^2}  \right) \nn \\
    \Gamma(N \to h\nu_\alpha)
    &\approx \frac{G_F M_N^3}{8\pi\sqrt{2}}  |\Theta_\alpha|^2 \left( 1 - \frac{M_h^2}{M_N^2} \right)^2\\
    \Gamma(N \to Z\nu_\alpha)
    &\approx \frac{G_F M_N^3}{8\pi\sqrt{2}}  |\Theta_\alpha|^2 \left( 1 - \frac{M_Z^2}{M_N^2} \right)^2 \left( 1 + \frac{2M_Z^2}{M_N^2} \right)\,,\nn
\end{align}
where both particle and antiparticle final states are taken into account.

For $M_N \gg M_{W,Z,h}$, the HNL lifetime~$\tau_N$ can be approximated as:
\begin{align}
\label{eq:HNL_lifetime}
     \nonumber\tau_N^{-1} &\approx \sum_{\alpha} \left[ \Gamma(N \to W\ell_\alpha)+ \Gamma(N \to Z\nu_\alpha) +\Gamma(N \to h\nu_\alpha) \right]\\
   &\cong \frac{ G_F}{2\pi\sqrt{2}} M_N^3 |\Theta|^2\,,
\end{align}
where the total mixing angle has been defined as $|\Theta|^2 = \sum_{\alpha=e,\mu,\tau} |\Theta_{\alpha}|^2$.

\begin{figure}%[tbh]
    \centering
    \includegraphics[width=.95\linewidth]{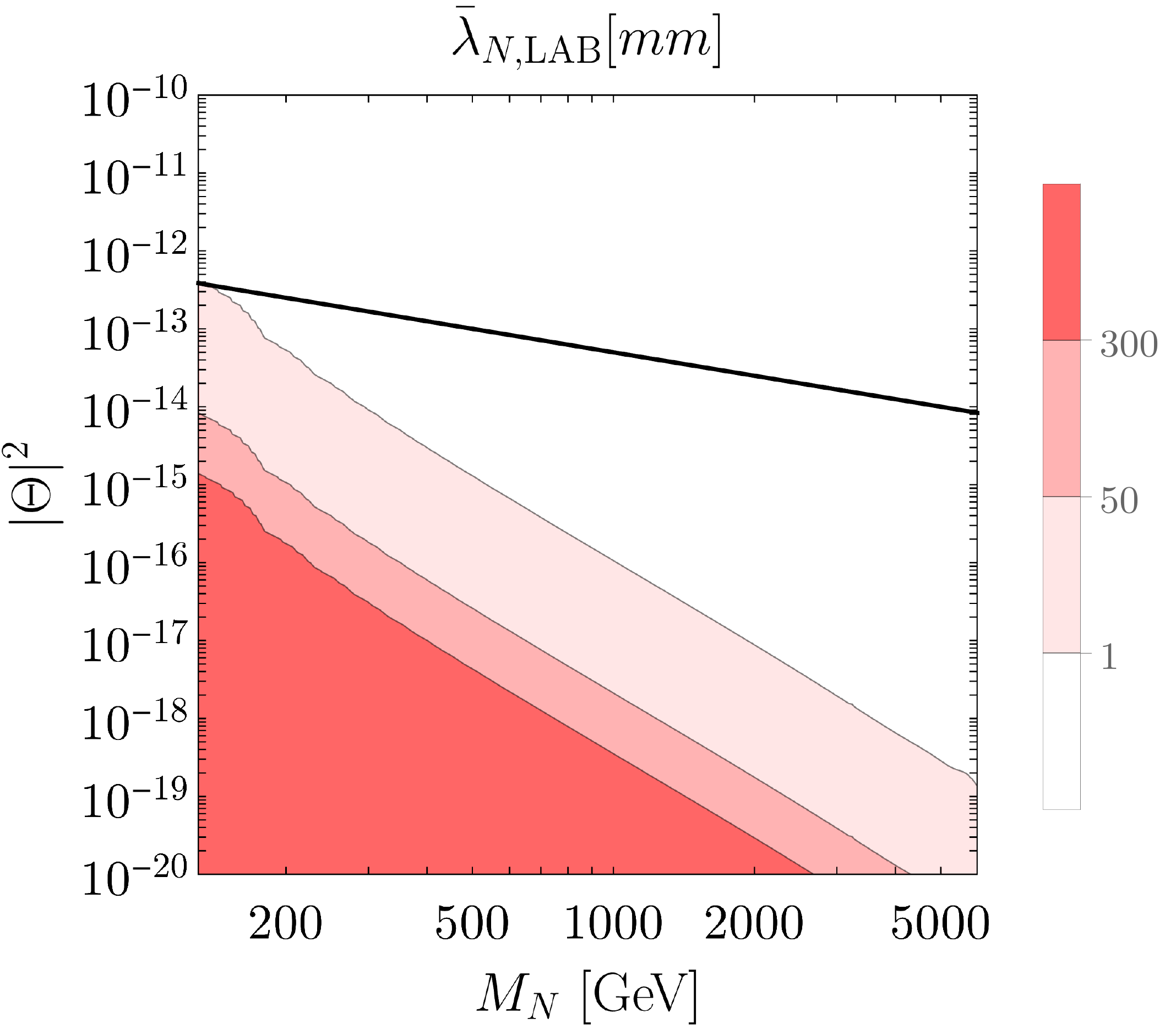}
    \caption{\em Contours of constant mean path of an HNL in the LAB frame as a function of $M_N$ and of the total mixing angle. The black line indicates the naive Type-I Seesaw expectation $|\Theta|^2=m_\nu/M_N$ with $m_\nu=50\,\mathrm{meV}$.
    The contours correspond to typical scales in the ATLAS experiment, with $1\,\mathrm{mm}$ being a typical cut in a prompt search, the interval $[50,300]\,\mathrm{mm}$ is covered by large radius tracking~\cite{ATLAS:2017zsd} while $[1,50]\,\mathrm{mm}$ is best covered by a combination of standard and large radius tracking.}
    \label{fig:lifetime}
\end{figure}

The mean path of an on-shell HNL in the LAB frame largely affects the ability of a search to detect it. In what follows we perform a rough estimation of this quantity.
The mean distance in the LAB frame $\lambda_{N,\text{LAB}}$, is given by
\begin{equation}
\lambda_{N,\text{LAB}}=\beta_N\gamma_N\tau_N=\sqrt{\dfrac{\hat{s}}{4M_N^2}-1}\,\tau_N\,,
\end{equation}
where we have taken into account the fraction of energy delivered by the partons to the HNL. Averaging over this quantity:
\begin{equation}
\ov{\lambda}_{N,\text{LAB}}=\dfrac{\iint_{\sqrt{\hat{s}}\geq 2M_N}\text{d}x_1\text{d}x_2 f_{g}(x_1)f_{g}
(x_2)\lambda_{N,\text{LAB}}}{\iint_{\sqrt{\hat{s}}\geq 2M_N}\text{d}x_1\text{d}x_2 f_{g}(x_1)f_{g}
(x_2)}.
\end{equation}
The result can be seen in Fig.~\ref{fig:lifetime}.
%^^^^^^^^^^^^^^^^^^^^^^^^^
\section{Event selection}
\label{app:efficiencies}
The projected sensitivity presented in this paper has been derived under the optimistic assumption of perfect acceptance. In this appendix, we test this assumption by applying a simple but typical selection (inspired by various type-III Seesaw and vector-like lepton searches that looked for similar topologies \cite{ATLAS:2015fma,ATLAS:2015qoy,CMS:2019hsm,CMS:2019lwf,ATLAS:2020wop,ATLAS:2022yhd}) and by explicitly computing the signal efficiency.
The leading lepton (defined as the lepton with the largest transverse momentum \pT{}) is required to have $\pT > \SI{40}{GeV}$, and the subleading lepton to have $\pT > \SI{15}{GeV}$. Both leptons are required to be within the pseudorapidity range $|\eta| < 2.5$ and to be separated from each other by $\Delta R > 0.4$. If the lepton pair has the same flavour but opposite signs (OSSF), its invariant mass is further required to be $m_{l^+l^-} > \SI{110}{GeV}$ to suppress background from $Z$ decays.
Jets (here quarks since we are performing the analysis at parton level) are required to have $\pT > \SI{30}{GeV}$ and to be within $|\eta| < 2.4$. We additionally require a separation $\Delta R > \Delta R_{jj}$ between any two jets (with a varying value of $\Delta R_{jj} = 0.001$, $0.2$ or $0.4$ in order to study its impact), as well as a separation of $\Delta R > 0.4$ between any given jet and lepton.

\begin{figure}[ht]
    \centering
    \includegraphics[width=\linewidth]{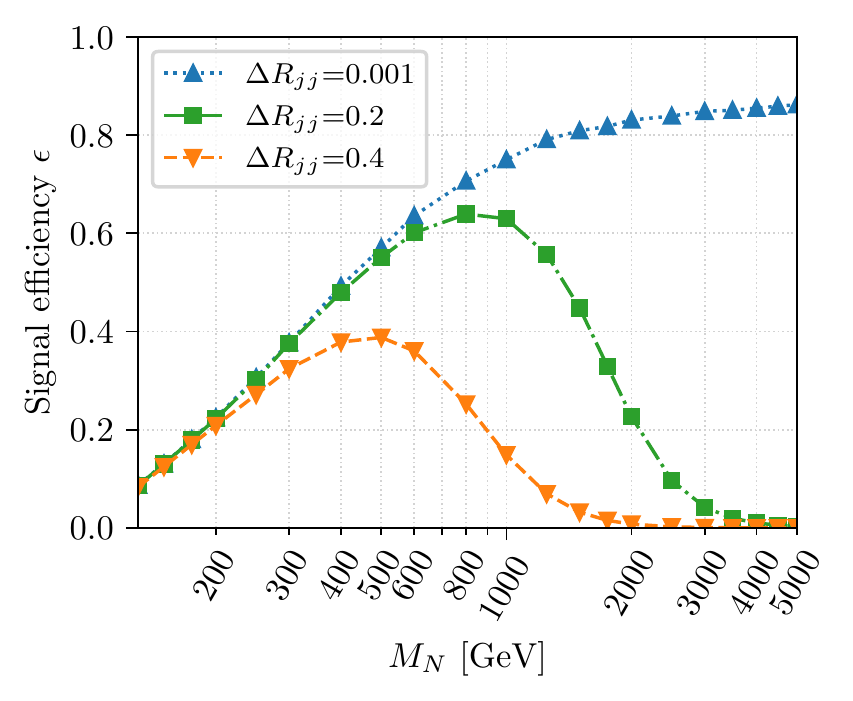}
    \caption{Signal efficiencies for the simple selection, as a function of the HNL mass $M_N$ and the minimum distance $\Delta R_{jj}$ between any two jets.}
    \label{fig:efficiencies}
\end{figure}

The cuts are applied directly in MadGraph's \texttt{param\_card.dat}, before generating the process \texttt{g g > n1 n1, n1 > j j l} with \texttt{l} denoting $e^{\pm}$ or $\mu^{\pm}$ but not $\tau^{\pm}$ (since the tau lepton efficiency has been treated separately) and \texttt{n1} denoting the HNL. We assume equal mixing of the HNL with all three flavours.
The signal efficiency, computed as the ratio of the generated parton-level cross-sections with/without cuts, is plotted in Fig.~\ref{fig:efficiencies} as a function of the HNL mass and $\Delta R_{jj}$.

We observe that imposing a non-trivial $\Delta R$ separation between jets strongly suppresses the signal at higher HNL masses, while other cuts are subleading. In practice, because small-$R$ jets are typically reconstructed with an angular size of $R=0.4$, for two jets to be resolved the parent partons will need to be separated by $\Delta R_{jj} \gtrsim 0.4$.
The suppression then has a simple interpretation: the partons produced in the decay of a boosted~$W$ are collimated in the LAB frame and form a single boosted jet instead of two resolved jets. Therefore, in order to be sensitive to the JALZ process at large HNL masses, one should aim to reconstruct the hadronic decay products of each $W$ as a single large-radius jet (with $R \approx 1$).
Performing an analysis at the reconstructed level is outside the scope of the present letter, and is left for further study. For the time being, we shall assume that the two large-$R$, $W$-tagged jets can be successfully reconstructed with $\mathcal{O}(1)$ efficiency.

Following a similar analysis and enabling each cut one-by-one, we observe that the cuts responsible for the suppression at lower HNL masses are, in rough order from the most to least important: the \pT{} cuts on the jets, the minimum $\Delta R$ between a lepton and jet, and the \pT{} cut on the leading lepton. Then comes the $Z$~veto and the $|\eta|$ requirements with a much smaller effect, while the remaining cuts are subleading.
The \pT{} cuts largely depend on the specifics of the detector and trigger system and should be a prime target for optimisation by the search team. Regarding the minimum $\Delta R$ between a lepton and a jet, it might be possible to do away with this cut altogether by once again trying to reconstruct the $W$ decays as large-radius jets possibly containing a lepton track.
While the \pT{} cuts could already have been implemented in the present study, we shall leave the calculation of precise signal efficiencies to an upcoming study.

\end{appendix}
%%%%%%%%%%%%%%%%%%%%%%%%%  Bibliography    %%%%%%%%%%%%%%%%%%%%%%%%
%%%%%

\footnotesize

\bibliography{bibliography}{}
\bibliographystyle{BiblioStyle}

\end{document}